\documentclass[letter]{aa}
\usepackage{graphicx}
\usepackage{txfonts}
\begin{document}

   \title{Detection of HNC and tentative detection of CN at $z=3.9$}
\titlerunning{Detection of HNC and CN at $z=3.9$}
\authorrunning{Gu\'elin, Salom\'e, Neri, Garc\'{\i}a-Burillo et al.}
   \author{M. Gu\'elin\inst{1}, P. Salom\'e\inst{1}, R. Neri\inst{1}, S.
Garc\'{\i}a-Burillo\inst{2}, J. Graci\'a-Carpio\inst{2},
J. Cernicharo\inst{3}, P. Cox\inst{1}, P. Planesas\inst{2},
P.M.~Solomon\inst{4}, L.J. Tacconi\inst{5} and P. Vanden
Bout\inst{6}}

   \offprints{M. Gu\'elin}

   \institute{IRAM, Domaine Universitaire, 300 rue de la piscine,
St Martin d'H\`eres F-38400, France.
              \email{guelin@iram.fr}
   \and
             Observatorio Astron\'omico Nacional, Calle Alfonso XII 3,
         E-28014 Madrid, Spain.
    \and IEM-DAMIR, CSIC, Serrano 121, E28006, Spain.
\and Dept. of Physics and Astronomy, State Univ. of N.Y., Stony Brook,
NY 11974, USA
\and Max-Planck-Institut f\"ur extraterrestrische Physik, Postfach
1312, 85741 Garching, Germany.
\and NRAO, 520 Edgemont road, Charlottesville, VA 22903, USA.}

   \date{Received October 13, 2006; accepted December 07, 2006}

\abstract {} {Molecular line emission from high-redshift galaxies holds great promise for
the study of galaxy formation and evolution.}{The weak signals can only be detected with
the largest mm-wave telescopes, such as the IRAM interferometer.} {We report the detection
of the J = 5--4 line of HNC and the tentative detection of the N= 4--3 line of CN in the
quasar APM~08279+5255 at $z=$3.9. These are the 4$^{\rm th}$ and 5$^{\rm th}$ molecular
species detected at such a high redshift.  The derived HNC and CN
line intensities are 0.6 and 0.4 times that of HCN J= 5--4. If
HNC and HCN are co-spatial and if their J= 5--4 lines are collisionally
  excited, the [HNC]/[HCN] abundance ratio must be equal to 0.6
within a factor of 2, similar to its value in the cold Galactic clouds and much larger than in the hot
molecular gas associated with Galactic HII regions. It is possible, however, that
fluorescent infrared radiation plays an important role in the excitation of HNC and HCN.}
{} {}


   \maketitle
%

\section{Introduction}

The presence of large reservoirs of molecular gas in the early Universe has been
demonstrated through the detection of rotational transitions of CO in high redshift
ultraluminous galaxies and quasars (see Solomon \& Vanden Bout, 2005, for a review). The
derived masses are in excess of $\rm 10^{10} \, M_\odot$ and the gas is found to be warm
and dense. Obviously,
a prodigious star formation activity is taking place in some of those objects, as attested
by the huge far-infrared luminosities. These considerations have triggered searches for
molecular species having higher dipole moments than CO and that are better probes of the
very dense gas associated with star formation. Two such molecules were detected so far in
high-$z$ sources: HCN and HCO$^+$.

The gravitationally lensed quasar APM~08279+5255 ($z=$3.9118, Wei\ss~ et al. 2006 [We06]) is
a prime target for such studies. Its huge intrinsic luminosity, boosted by a large
magnifying factor ($m$=60-100), makes it not only the most luminous object in the Universe
(apparent luminosity $L_{bol}=7$ 10$^{15} L_\odot$), but also one of the most powerful
sources of CO emission. The presence of strong CO lines with rotational quantum numbers as
high as J = 11--10 indicates that its gas is very dense and warm (Downes et al. 1999,
We06). Both HCN and HCO$^+$ have been detected in this source (Wagg et al. 2005 [Wa05];
Garc\'{i}a-Burillo et al. 2006 [GB06]).

To further constrain the physical conditions of the gas in APM~08279+5255 and to probe its
chemical composition, we have searched for new high density tracers. In this Letter, we
report the first detection of hydrogen isocyanide (HNC) at high-$z$ and the tentative
detection of CN.

\begin{figure}
\centering
\includegraphics[width=6cm, angle=-90]{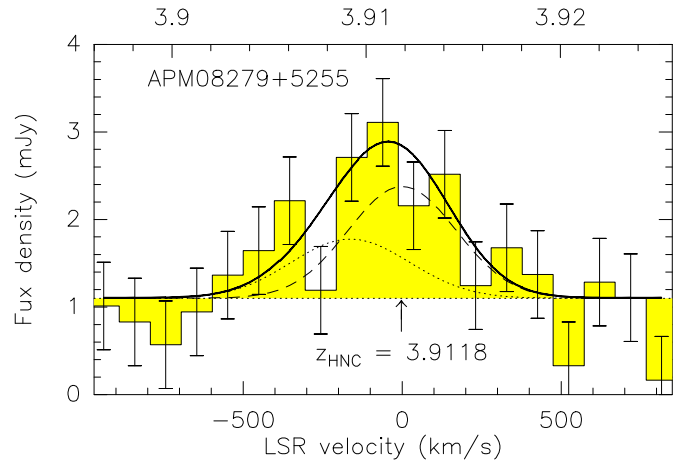} 
\caption{Spectrum of the HNC(5--4) and CN (4--3) emissions from APM~08279+5255,
The velocity scale is relative to the HNC frequency redshifted by $z=3.9118$. 
The velocity resolution is 97 km s$^{-1}$, the r.m.s. noise 0.5
mJy. The thick line represents the best fit synthetic spectrum and
the dotted and dashed lines the contributions from CN and HNC to this
spectrum. The CN contribution is the blend of 2 fine-structure components.} \label{spec}
\end{figure}

\section {Observations and Results}

The IRAM Plateau de Bure interferometer (PdBI) was used to observe the HNC ${\rm J=}$5--4
and CN ${\rm N=}$4--3 lines which, redshifted by $z=3.9118$ to the 3-mm band, have rest
frequencies differing by less than 70 MHz. More specifically, HNC ${\rm J=}$5--4 is a
single component line with a redshifted frequency of 92.282~GHz, while CN ${\rm N=}$4--3,
due to spin-coupling and spin-spin interactions, shows a
complex line pattern, whose strongest components lie close to 92307 MHz and 92350 MHz and have
intensities in the ratio 1:1.3.

The observations were made in September 2006 with five antennas in a compact configuration.
The receivers were tuned to 92.297~GHz. The SSB system temperature was typically
$150-200$~K. The 560~MHz-wide IF-bandwidth, covering a velocity range of 1820~km~$\rm
s^{-1}$, was observed with a channel spacing of 2.5~MHz. The FWHP of the synthesized beam
is 5.6$^{\prime\prime}$ $\times$ 4.4$^{\prime\prime}$. Data reduction and analysis were
done with the GILDAS software.  The flux calibration is based on the PdBI primary
calibrator MWC~349 and on reference quasars. The total time on source amounts to 25 hours.

The resulting spectrum, smoothed to a resolution of 97 kms$^{-1}$, is shown in Fig. 1. The
total emission map, integrated over the 1820 kms$^{-1}$-wide bandwidth, is shown in Fig.
2a, while 3 continuum-free velocity-channel maps, obtained after subtraction of a 1.1 mJy
continuum source, are shown in Fig. 2 (b-d). A line is clearly detected on Fig. 1 and Fig.
2c. Like the continuum emission, it arises from an unresolved source whose position is
identical to those of the CO, HCO$^+$ and HCN sources and of the quasar. The line peak
flux, derived from the fit of a single Gaussian profile, is 1.76 $\pm\, 0.3 $ mJy ($6
\sigma$ detection). The line width is 500 $\pm$ 110 km s$^{-1}$ and its center frequency,
relative to the LSR, is $92294 \pm 12$ MHz. The latter is 12 MHz higher than the HNC line
frequency and 38 MHz lower than that of the barycenter of the CN line components. The
fitted width and center frequency, which are somewhat larger than those observed for HCN,
suggest that we have detected a blend of HNC and CN.

We have fitted the unsmoothed line profile with a synthetic profile representing a blend of
the HNC and CN lines. The relative position of the line components (2
components for CN, one for HNC) were fixed according to
the redshifted transition frequencies and the component widths were
set equal to $\Delta {\rm v}=
400$ kms$^{-1}$, the FWHP of the HCN (5-4) line(We06). The intensities derived for
$z=3.9118$, the redshift derived from CO, are shown in Table 1. They correspond to a HNC/CN
integrated line intensity ratio of $r=1.7$. Obviously, CN is only a minor constituent of
the observed line. The observed profile can be marginally fit by HNC alone (1 $\sigma$),
but not by CN alone: fitting this profile with only the CN components yields a much too
high redshift, $z= 3.9139 \pm 0.0008$. We conclude that the detection of HNC in
APM~08279+5255 is certain, whereas that of CN is only tentative.  A fit of the component
width, while keeping the intensities fixed, yields $\Delta v= 480 \pm 100$ kms$^{-1}$ a
value similar, within the errors, to the HCN and HCO$^+$ line
widths. This  is consistent with
HNC, HCN and HCO$^+$ being co-spatial.

The derived velocity-integrated HNC and CN line intensities (and the corresponding line
luminosities) are compared in Table~1 to those of HCN, HCO$^+$ (5--4) and CO
(4--3). The relatively large error bars reflect the difficulty of resolving CN
from HNC. The HNC line intensity is surprisingly large: $\simeq 2/3$ of those of HCN and
HCO$^+$ and $\simeq 1/7$ of that of CO. The CN line intensity is
certainly smaller and remains very uncertain.

\section{Discussion}

The detection of HNC and the tentative detection of CN in APM 08279+5255, after those of
CO, HCN and HCO$^+$, brings to 5 the number of molecules observed in high redshift quasars.
More than just supplying new tracers of dense gas, it gives us an opportunity to measure
the [HNC]/[HCN] abundance ratio, a ratio very sensitive to the physical and chemical state
of the gas.

The strength of HNC, HCN and HCO$^+$ in APM 08279+5255 may reflect a high
abundance of these species relative to CO. It may also come from an unusually high gas
density. Finally, it may result from fluorescent pumping through excited bending states, or
from a combination of those three causes (see GB06).

Whereas the abundance of HCN, relative to CO, is stable in Galactic clouds for a wide range
physical conditions (Lucas \& Liszt 1996), that of HNC is known to
vary by orders of magnitude. The [HNC]/[HCN] ratio is found to be $\geq 1$ in dense dark
clouds (Hirota et al. 1998), $\simeq 1/5$ in diffuse clouds (Liszt \& Lucas 2001) and only
few $\times 10^{-2}$ in hot and dense star-forming regions, such as the Orion hot core
(Schilke et al. 1992).

In dense clouds, both HCN and HNC are thought to mainly result from the dissociative
recombination of HCNH$^+$ and to be destroyed by reactions with ions
and radicals. HNC 
is preferentially destroyed by reactions with H, O and OH that proceed only
in oxygen-rich hot and/or dense environments. Schilke et al. (1992) have modeled the
[HNC]/[HCN] abundance ratio to explain the Orion results. They predict a rapid decrease of
this ratio with increasing gas density and temperature, in agreement with observations: in
their model, [HNC]/[HCN] decreases from a value of $\geq 1$ at $n
<10^5$~cm$^{-3}$, $T_K<30$~K (the conditions prevailing in dense dark clouds) to
[HNC]/[HCN] $\sim 3 \, 10^{-2}$ at $n= 10^6$~cm$^{-3}$ and $T_K= 100 $~K (the conditions in
the Orion hot, oxygen-rich core).

In the following, we examine the physical conditions prevailing in the circumnuclear disk
of APM~08279+5255 and address the question of HNC excitation, in order to discuss the [HNC]/[HCN]
abundance ratio.

\begin{table*}
\caption{Properties of the HNC, HCN and HCO$^+$ (5--4) lines compared to the CO(4--3) line towards APM~08279+5255}
\begin{center}
\begin{tabular}{lccccc}
\hline
\hline
Line             & $\rm \nu_{obs}$  & I$_{line}$         & L$^\prime$                  & F$_{cont}$    & Ref \\
                 &    [GHz]         & [Jy km s$^{-1}$]          &
[10$^{10}$K km s$^{-1}$ pc$^{2}$]  &    [mJy]      &     \\
\hline
HNC(5--4)        & 92.282           & 0.54$^d\pm$0.25  & 2.3$\pm$1.1 &          1.1$^e\pm$0.15 & Present work \\
CN(4--3)         & 92.332           & 0.31$^d\pm$0.25    &1.3 $\pm$1.1             &   1.1$^e\pm$0.15        & Present work \\
HCN(5--4)        & 90.229           & 0.85$\pm$0.12      & 3.6$\pm$0.5                 & 1.3$\pm$0.2   &   (a)  \\
HCO$^{+}$(5--4)  & 90.797           & 0.87$\pm$0.13      & 3.5$\pm$0.6                 & 1.2$\pm$0.1   &   (b)  \\
CO(4--3)         & 93.870           & 3.70$\pm$0.50      &14.7$\pm$1.5                  & 1.2$\pm$0.3   &   (c)  \\
\hline
\end{tabular}
\label{parameters}
\end{center}
 NOTES. -- (a) We06; (b) GB06; (c) Downes et al. (1999); (d) blended line
          HNC and
          CN fitted together with FWHP fixed
          to 400 km s$^{-1}$; (e) fitted to the line-free channels. 
          The luminosities are calculated from equation [1] of Downes et al. (1999) using the standard
concordance cosmology parameters with $\rm H_0 = 71 \, km s^{-1} Mpc^{-1}$,
$\Omega_m = 0.27$ and $\Omega_\Delta = 0.73$ (Spergel et al. 2003). They are not corrected for the lens
amplification.

\end{table*}

\begin{figure*}
\centering
\includegraphics[width=16cm]{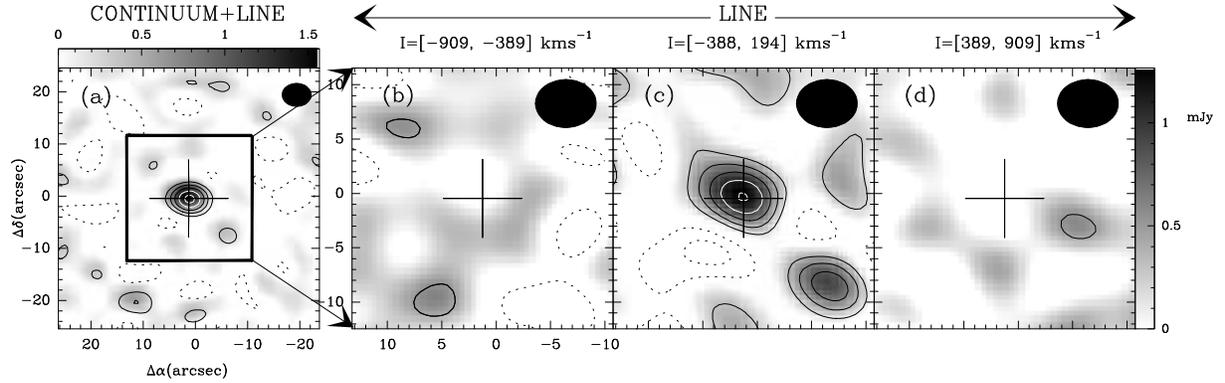}
\caption{{\bf a}: Total (continuum+line) emission map of APM~08279+5255 integrated from
$-909$ to $\rm 909 \, km s^{-1}$ and centered on 92.29~GHz. Levels are $-2 \sigma$ (dashed)
and $2 \sigma$ to 10 $\sigma$ in steps of 2 $\sigma$ ($\sigma$ = 0.15 mJy/beam). {\bf b-d}:
Velocity-channel maps obtained after subtraction of a point-like 1.1 mJy continuum source,
located at the position of maximum emission.  The data have been averaged over the three
velocity intervals delineated in Fig.~1. {\bf b:} $\rm I=[-909, -389]$, {\bf c:} $\rm
I=[-388, +194]$ and {\bf d:} $I=[389,909]\, km s^{-1}$. The emission from the HCN(5--4) and
CN (4--3) lines appears in the central channel ({\bf c}). Contours are $-2\sigma$,
$-1\sigma$ (dashed) and 1$\sigma$ to 5$\sigma$ in steps of $\sigma$ for (c) and $-2\sigma$
and $2\sigma$ for (b) and (d) (with $1\sigma$ = 0.25 mJy/beam).  The synthetized beam is
shown as a grey ellipse on the top right corner. The offsets (in arcseconds) are relative
to the peak of the continuum+line emission at ($\rm 08^h31^m41^s69, 52^o45'17''12$)
(J2000), shown by the crosses.
}
\label{cont_line}
\end{figure*}

\subsection{Collisional excitation}

The physical conditions in APM~08279+5255 can be derived, in principle, from the observed
CO and HCN line intensities. CO and HCN have different dipole moments and their rotational
transitions have critical densities in the ratio 1:1000, making the HCN/CO intensity ratio
a sensitive indicator of the gas density. Downes et al. (1999), Wa05 and We06 have
analyzed, using a LVG code, the CO J=1--0 through 11--10 and HCN(5--4) line intensities in
APM~08279+5255.  They assume that the observed emission arises from a single uniform
component, or from two radiatively decoupled components, and that fluorescent excitation is
negligible. With these assumptions, the range of densities for the bulk of the gas
is constrained by the fact that the gas density $n$ must be high enough to populate the J =
5 level of HCN and low enough to prevent the thermalization of the CO J $\geq$ 9 levels.
This range depends mildly on the gas temperature (T$_K$) or the molecular abundances and can
be derived for reasonable values of those parameters, i.e. ${\rm 40 K < T_K < 200}$ K, $\rm
10^{-5} < [CO]/[H_2] < 10^{-4}$ and ${ 10^{-4} <}$ [HCN]/[CO] ${\rm < 10^{-2}}$). The gas
density $n$ should be comprised between a few $\times 10^4 {\rm cm^{-3}}$ and a few $\times
10^5 {\rm cm^{-3}}$. The solution favored by We06, who assumed [CO]/[H$_2$]= $5 \,10^{-5}$
and [HCN]/[CO]=$10^{-3}$, is $n= 10^5 {\rm cm^{-3}}$, $T_K=45$ K; Wa05, who adopted
N(CO)/$\Delta$v= $4\, 10^{17}$ cm$^{-2}$km$^{-1}$s and [HCN]/[CO]= $10^{-2}$, find $n= 4\, 10^4 {\rm cm^{-3}}$ and $T_K=80$ K. The HCN(5--4)
line opacities ($\tau$) derived in these studies are 50--150.

With such values of $n$ and $\tau$, the HCN(5--4) line, which has a critical density of $2
\times 10^7$~cm$^{-3}$, is only weakly excited. Low excitation implies that the intensity
emerging from a uniform cloud is nearly proportional to the molecule column density N(HCN).
As an illustration of this, let us consider the solution with $n= 10^5
{\rm cm^{-3}}$,  $T_K=45$ K and a line opacity $\tau=100$. A
statistical equilibrium calculation with
RADEX  (Schr\"{o}ier et al. 2005) shows
that dividing N(HCN) by 3, decreases the HCN(5--4) line intensity by a
factor of 2.5, despite the large value of $\tau$. The same considerations apply to HNC, which has almost
exactly the same electric dipole moment as HCN (3.0 debye) and similar rotational level
spacings. Recent calculations by Wernli et al. (2006) for the collisional excitation of
HCCCN with He and with H$_2$ and on-going calculations for HCN, based upon recent potential
energy surfaces (M. Wernli \& P. Valiron, {\it private communication}), indicate that for
such heavy molecular rods, the dynamics of the collision is mainly driven by the hard core
interaction with the rod and is weakly sensitive to the details of the potential surface,
in fair agreement with the pioneering calculations for HCN-He by Green \& Chapman (1978).
We can thus safely assume that HNC and HCN present similar collisional rates and critical
densities for temperatures well below the isomerisation barrier of 1440 K. Thanks to this
and provided HCN and HNC co-exist in the same clouds, the HNC/HCN intensity ratio in
APM~08279+5255 should reflect these species' abundance ratio. In that case,
[HNC]/[HCN]$\simeq 0.6$ within a factor of 2, close to the ratio predicted by Schilke et al. (1992) in O-rich environments for
$T_K\leq 80$ K, $n\leq 10^5$ cm$^{-3}$, but higher than those
predicted for hotter and denser regions, such as the Orion hot core.

The [CN]/[HCN] abundance ratio is more difficult to derive for 3 reasons. First, the CN line intensity
is fairly uncertain. Second, the $\rm J =$ 4--3 CN line is much easier to excite than HCN
(5--4), due to its lower energy and smaller dipole moment, so that the
line intensity ratio depends critically on the
gas density. Third, the collisional cross sections of CN with H$_2$, or even He, are not
known. Following Black \& van Dishoeck (1991), we assume that they are similar to those of
CS (Green \& Chapman 1978) and adopting once more $T_K=45$ K,
$n=10^5$ cm$^{-3}$, we estimate [CN]/[HCN]$\leq 1/10$.

\subsection {Fluorescent excitation}

We have assumed above that there is no radiative coupling between the
different clouds in the ${\rm J=}$5-4 line and that the excitation is
mainly collisional. The  first  hypothesis seems reasonable in view of
the limited inclination (the quasar is bright at visible wavelengths)
and large velocity gradient of the rotating disk. The fact that we are
dealing with J levels as large as 4 and 5 makes it also unlikely that
cold foreground gas screens
efficiently the core emission. The second hypothesis is weaker.

With a spectral energy distribution peaking around 20$\rm \mu m$,
APM~08279+5255 is exceptionally bright in the mid-infrared.
A  fit to the dust emission yields a hot (200~K) plus a cold (70~K)
component (Beelen et al. 2006).  The latter, whose mass represents
more than 9/10 of the total, is probably associated with the dense gas
detected in the HCN and HNC lines. The hot dust component could be
concentrated close to the central AGN (We06), or distributed in hot
flakes scattered throughout the cold disk (Nenkova et al. 2002).

The transitions  $\nu_2$=1--0 connecting the lowest excited bending
states of HCO$^+$, HCN and HNC to the ground vibrational state
have wavelengths of 12.1, 14.0 and 21.7$\mu$m, respectively. Their Einstein
A coefficients are fairly large
($A\simeq 1-7 \, {\rm s^{-1}}$, Nezu et al. 1999),
so that the $\nu_2=$1--0 lines are
expected to be optically thick in APM~08279+5255.  As argued by GB06
(see also Barvainis et al. 1997),
the mid-infrared radiation from the hot dust can well excite the  first
bending modes and, by fluorescence, populate the J= 4 and 5 levels of
the ground state. Since the J=3 levels are already populated by the
cosmic background radiation (their fractional population is 0.17 at
13.4~K), one single pumping cycle -- from the ($\nu_2$,J)= (0,3) level
up to the (1,4) level and down to the (0,5) level-- is required to
populate the J = 5 level. Besides an optical depth $\geq 1$, the only
requirement for fluorescent pumping to be efficient is that the solid
angle $\Omega$ subtended by the hot dust at the molecules should not
be too small (say $f=\Omega/4\pi\geq 0.1$). A value of $f> 0.1 $ could
be easily achieved if the gas is clumpy, so that the radiation from
the AGN can penetrate deeply into the disk (Nenkova et al. 2002).

If those conditions are fulfilled, fluorescent excitation may well take over collisional
excitation for the HCN, HNC and HCO$^+$ J = 5--4 lines. The line intensities may then
reflect the number of pumping photons, rather than the molecular column densities. The near
equality of the HCN and HCO$^+$ J= 5--4 line intensities would be naturally explained, as
the infrared transition wavelengths of these two species are similar. Also,  the constraint
on the gas density would be much relaxed, since the CO data alone can be explained
by warmer, but less dense gas (We06).

Infrared pumping may introduce interesting differences between HCN and
HNC. The wavelength of the $\nu_2=$1--0
transition of HNC is 1.5 times larger than that of HCN and the flux of APM~08279+5255
at 21.7~$\rm \mu m$ is twice that at $14 \mu$m. Thus, the
number of photons able to excite
the HNC molecules is at least 3 times larger than that of photons which can
excite HCN (or HCO$^+$). Moreover, the emission of the colder dust
component starts to be  significant at 21.7~$\mu$m. This emission will be
particularly effective, since the cold dust is better mixed
with the HNC molecules than the hot dust, so that $f\simeq 1$.

In contrast to triatomic molecules that have low energy bending states, the lowest
vibrational transition of CN lies at 4.9 $\mu$m, where the flux of
APM~08279+5255 is 5 times weaker than at 20$\mu$m. The comparison of
HCN, HNC and CN may then offer a way to weigh the
relative importance of collisional and fluorescent excitations, assuming these molecules
are co-spatial. More transitions would be needed, however, to do so.


\subsection{Comparison with other galaxies}

In nearby starbursts galaxies such as M~82, NGC~253, NGC~1068 and
NGC~3079, the HNC and HCN ${\rm J=}$1-0 lines have intensity ratios
$\simeq 0.5$ (H\"uttemeister et al. 1995; Wang et al. 2004). Exceptions are
some ULIRGs such as Mrk~231 (Aalto et al. 2002) and Arp 220 (Cernicharo
et al. 2006), where this ratio is $\simeq 1$. However, the fundamental
rotational lines are not necessarily good indicators of the HCN and
HNC abundances, due to self-absorption. First, these lines have
critical densities of 2\, 10$^5$ cm$^{-3}$, 100 times lower than the
$\rm J=$ 5--4 lines; they are much easier to excite in dense cores,
so that their intensities saturate for lower molecular column
densities.  Second, in the local Universe the cosmic background
temperature is only 2.7 K, so that most of the HNC and HCN molecules
in low density gas are in the ground ${\rm J=0}$ level, making
envelopes optically thick to the ${\rm J=1-0}$ line radiation that
emerges from the cores.

A comparison between local ULIRGs and APM 08279 +5255 must involve
higher J lines. The ${\rm J=3-2}$ HNC and HCN lines have recently been
observed in Arp~220 (Cernicharo et al. 2006) with an intensity ratio
of 2.3, twice the value of the $\rm J=$1--0 line intensity ratio. The
J=3--2 line intensity ratio is more likely to reflect the [HNC]/[HCN]
abundance ratio than the $\rm J=$1--0 intensity ratio, so that HNC
should be more abundant than HCN in this source. The relatively high
HNC abundance observed in APM~08279+5255 is thus not
exceptional.

A survey of CN (N= 1--0) emission in luminous IR galaxies has been made by Aalto et al.
(2002), who found that the CN/HCN intensity ratio, which is $\simeq 1$ in many LIRGs  can
be lower in ULIRGs. The relative weakness of CN in APM~08279+5255 is therefore
also not exceptional.

\section{Summary}

We have detected HNC (5-4) and tentatively CN (4-3) emission from the quasar
APM~08279+5255 at $z=3.9$, adding to HCN and HCO$^+$ two new tracers of
the very dense gas in high-$z$ sources. The data are consistent with HNC and
HCN being co-spatial.

The $\rm J=5-4$ lines of HCN and HNC are remarkable by their very high, almost equal
critical densities, $\simeq 2 \, 10^7$ cm$^{-3}$, which are far larger than the gas density
in the central circumnuclear disk. In the absence of mid-infrared pumping, the high
critical densities maintain low populations in the $\rm J = 5$ and 4 levels. The
HCN/HNC(5--4) line intensity ratio is then a good measure of the [HNC]/[HCN] abundance
ratio. The latter, which is a sensitive probe of the chemical and physical conditions, is
found to be close to 1, much larger than the corresponding ratio observed in the 
hot and dense Galactic molecular clouds. 
We stress that the $\rm J=1-0$ lines of the main isotopologues of
HNC and HCN are much more easily excited and do not trace properly these species'
abundances when they are optically thick.

HNC is the first metastable isomer detected in high-$z$ sources.
CN, if confirmed, would be the first radical. Their observation in APM~08279+5255
illustrates that the chemistry can be quite evolved in environments as extreme as the
vicinity of the most powerful high-$z$ quasars. Other molecules, including more complex
species, might be detectable with present day instrumentation.

\acknowledgements{ We wish to thank P. Valiron for helpful comments on
 the collisional cross sections of HNC and HCN and A. {Wei\ss} and
 D. Downes for communicating their results prior to publication. IRAM is supported by
 INSU/CNRS (France), MPG (Germany) and IGN (Spain).}

\section{References}
Aalto, S., Polatidis, A.G., H\"uttemeister, S., Curran, S.J. 2002,
A\&A 381, 783\\ 
 Barvainis, R., Maloney, P., Antonucci, R., Alloin, D. 1997, ApJ 484, 695 \\
Black,
J., van Dishoeck, E.F. 1999, ApJ 369, L9\\ Beelen, A., Cox, P., Benford, D.J. et al. 2006,
ApJ, 642, 694 \\ Cernicharo, J., Pardo, J. R., Wei\ss, A. 2006,
ApJ. 646, L49 \\ 
Downes, D., Neri, R.,Wiklind, T.
et al. 1999, ApJ, 513, L1 \\
Garc\'{\i}a-Burillo, S., Graci\'a-Carpio, J., Gu\'elin, M., et al. 2006, ApJ 645, L 17 [GB06]\\
Green, S., Chapman, S. 1978, ApJS 37, 169\\
Hirota, T., Yamamoto, S., Mikami, H., Ohishi, M. 1998, ApJ 503, 717\\
H\"uttemeister, S., Henkel, C., Mauersberger, R. et al. 1995,  A\&A 295, 571 \\
Liszt, H., Lucas, R. 2001, A\&A 370, 576\\
Lucas, R. Liszt, H. 1996A\&A 307, 237\\
Nenkova, M., Ivezic, Z., Elitzur, M. 2002, ApJ 570, L9\\
Nezu et al. 1999, J. Mol. Spectrosc. 198, 186\\
Schilke, P., Walmsley, C.M., Pineau des For\^ets, G. et al. 1992, A\&A
256, 595\\
 Schr\"{o}ier, F.L., van der Tak, F.F.S., van Dishoeck E.F., Black, J.H. 2005, A\&A 432,
369-379 \\
Solomon, P.M. \& Vanden Bout, P.A. 2005, ARAA, 43, 677 \\
Spergel, D.N. et al. 2003, ApJS, 148, 175 \\
Wang, M., Henkel, C., Chin, Y.-N. et al. 2004, , A\&A 422, 883 \\
Wagg, J., Wilner, D.J., Neri, R.et al. 2005, ApJ, 634, L13 [Wa05]\\
Wei\ss, A., Downes, D., Neri, R. et al. 2006, A\&A {\it submitted} [We06]\\
Wernli, M., Wiesenfeld, L., Faure, A., Valiron, P. 2006, A\&A {\it submitted}

\end{document}